\documentclass[twocolumn,showpacs,epsfig,pre]{revtex4-1}
\usepackage{graphicx}
\usepackage{amssymb}
\usepackage{amsmath}

\usepackage{color,soul}

\newcommand{\Lop}{{\cal L}}

\newcommand{\pS} {{\cal M}}
\newcommand{\intM}[1]{{\int_\pS{\!d #1}\:}}
\newcommand{\prpgtr}[1]{\delta\negthinspace\left( {#1} \right)}
\newcommand\xInit{{x_0}}
\newcommand\flow[2]{{f^{#1}(#2)}}

\newcommand{\beq}{\begin{equation}}
\newcommand{\ee}[1] {\label{#1} \end{equation}}

\newcommand{\bea}{\begin{eqnarray}}
\newcommand{\eea}{\end{eqnarray}}

\begin{document}

%\draft
\title{Localization in chaotic systems with a single-channel opening}
\author{Domenico Lippolis$^{1,2}$\email{domenico@tsinghua.edu.cn}, Jung-Wan Ryu$^{2,3}$, and Sang Wook Kim$^4$}
\affiliation{$^1$Institute for Advanced Study, Tsinghua University, Beijing 100084, China}
\affiliation{$^2$Department of Physics, Pusan National University, Busan 609-735, South Korea}
\affiliation{$^3$School of Electronics Engineering, Kyungpook National University, Daegu 702-701, South Korea}
\affiliation{$^4$Department of Physics Education, Pusan National University, Busan 609-735, South Korea}
\date{\today}

\begin{abstract}
We introduce a single-channel opening in a random Hamiltonian and a
quantized chaotic map:
localization on the opening
occurs as a sensible deviation 
of the wavefunction statistics from the predictions
of random matrix theory, even in the semiclassical limit.
Increasing the coupling to the open channel in the quantum model, 
we observe a similar picture to resonance trapping, made of 
few fast-decaying states, whose left (right) eigenfunctions are entirely localized 
on the (preimage of the) opening, and  
plentiful long-lived states, whose probability density is instead suppressed at the
opening. 
For the latter we derive and test a linear relation
between the wavefunction intensities and the decay rates,
similar to Breit-Wigner law.
We then analyze     
the statistics of the eigenfunctions of the corresponding  (discretized)
classical propagator, finding a similar behavior to the quantum system only
in the weak-coupling regime.
\end{abstract}
% 05.70.Ln Quantum chaos
% 03.67.-a Localization
% 05.30.-d Opening
% 89.70.Cf Quantum maps
% 03.65.Ta Scars
% 05.70.-a 

\maketitle
%\narrowtext

%%%%%%%%%%%%%%%%%%%%%%%%%%%%%%%%%%%%%%%%%%%%%%%%%%%%%%%%%%%%%%%%%%%%%%%%%%%%%%%%%%%%%%%%%%%%%%%

\section{Introduction}

One of the distinctive traits of all chaotic 
systems is their seemingly `random' behavior \cite{EckRu}.
As a consequence,
one usually assumes that 
%of that is the validity of the
%so-called random wave assumption, namely that 
the eigenfunctions of a quantized chaotic Hamiltonian
have the same statistical properties (\textit{i.e.} 
wavefunction intensity distribution) of 
a complete set of waves with
random amplitudes and 
phases \cite{BerryTab,stoeck},
or equivalently, of the eigenvectors of a 
Hermitian matrix with random entries, according to 
random matrix theory (RMT) \cite{wigner,Mehta}.  
Due to a number of applications (quantum information theory \cite{Prosen,GeorgShep,Zurek},
classical \cite{noeck,Hentschel} and quantum optics \cite{AcMiShi,moore,embran}, 
quantum transport \cite{Benak,Miao}), as well as to equally many 
theoretical issues (see for example \cite{diff,Kap_opn,weyl}), the quantum chaos community is nowadays
largely focused on the behavior of \textit{open} systems \cite{Petruccio}. 

In this paper we address one of the simplest 
theoretical questions: 
whether and how the wavefunction statistics deviates from
the predictions of the random wave assumption as we 
%practice a small hole 
%in the phase space of a chaotic system.   
perturb a chaotic system with a single-channel 
opening. 
As main result of our investigation, we numerically find that 
 the overall wavefunction intensity distribution
at the location of the opening does change from 
the RMT-expected $\chi^2-$shape to a 
longer-tailed curve, which 
is analytically described
using perturbation theory.
It physically implies that localization occurs at the opening.
In our theory the opening can be an arbitrary state $|a\rangle$ 
in the Hilbert space, however, in most of our testing models we take it as a 
coherent state in the phase space.
Deviations of the wavefunction statistics from RMT have been observed before in real space: 
for time- reversally symmetric systems,
it was conjectured~\cite{PniShap}
and then shown analytically and experimentally~\cite{Sebaetal} that the distribution
of the wavefunctions at the leads
smoothly crosses over from Porter-Thomas' to Poisson's with the coupling
to the opening. 
Although there was no explicit mention of localization, the wavefunction distribution for a two-channel opening was found to be
an inverse square-root, of much slower decay than the RMT prediction. 
In a later work~\cite{Ishioetal}, this behavior was related to the correlations between real and imaginary parts of the wavefunction, which in general 
may depend on the underlying classical dynamics.

On the other hand, real- and phase-space localization
have been detected in closed systems 
in correspondence of the so-called
scars~\cite{Hel84,kap_hel}. Within that framework,
the distribution of the intensities on an unstable periodic orbit  
was found to decay
slower than the RMT-expected~\cite{kap_prl}, 
due to a phenomenon of constructive interference. 
This is not our case:
in order to rule out scarring, we place our probe states
away from periodic orbits. Still, the localization
found for weak coupling to the opening does hold in the
semiclassical limit, which makes us think of a classical
effect.
%Statistically,  
%the localization is ascribed instead to  
%the interaction of
%uncorrelated random eigenstates of
%the closed Hamiltonian 
%via the open channel.  
%The resulting eigenstates of the open Hamiltonian
%are statistically correlated,
%and give rise to a longer tail in 
%the distribution of the intensities.

Successively, we follow the evolution of the 
wavefunction statistics
of the quantum map for strong coupling to the opening.  
As a result, the intensity distribution becomes  
separated into several
long-lived- and a few short-lived
eigenstates. 
We show that their intensities are proportional
to their decay rates, 
arguing that this quantum effect 
can be explained with the existing theories on
resonance trapping~\cite{rotter91,russky}.
In particular, the intensities of the long-lived states
depend on the escape rates through a linear relation akin to   
Breit-Wigner law~\cite{stoeck}.     

In the second part of the paper we 
%address the same problem in a  
%classical setting,
%we may as well wonder whether the
%same intensity enhancement
%also occurs in a classical 
%since
%both the chaoticity of the system, and
%our opening (a hole in the phase space) are classical 
%ingredients, the deviation observed might
%not be originally quantum in nature.  
%fully-chaotic dynamical system
%with a small opening.  
%and
%For that reason we 
perform analogous simulations on the classical cat 
map and, 
by looking at the statistics of the
eigenfunctions of the classical propagator 
(Perron-Frobenius operator \cite{chaosbook}),
we find the deviation
from the closed system in all similar to 
the quantum case for weak coupling to the opening.
This observation corroborates the hypothesis of 
a classical mechanism behind localization, in this regime.
On the contrary, we show that a strongly-coupled  
opening does not result in resonance-trapping,
which makes the classical setting substantially different from the
quantum, in this regime.
%From that we suggest the 
%localization on the opening 
%might well be a classical effect in weak-coupling regime.

The paper is organized as follows: 
in section~\ref{sec:supertheory} we calculate the deviation 
of the wavefunction statistics from an exponential distribution
due to a single-channel opening by using 
first-order perturbation 
theory. In section \ref{sec:modello} 
we verify the theoretical expectation   
using random Hamiltonians drawn from the Gaussian unitary ensemble
(GUE)~\cite{Mehta}, and successively on the eigenfunctions of the quantized
cat map~\cite{creagh}. 
Section \ref{sec:three} 
deals with the strong-coupling regime:
%, together 
%with Husimi projections of the shortest-lived, most 
%localized left and right eigenfunctions. 
we analyze the proportionality between escape rates and intensities, 
%(section~\ref{sec:trap}),
while we
account for the 
localization patterns of left and right fastest-decaying eigenfunctions
in section~\ref{sec:obs}.
In section \ref{sec:classic} we introduce the Perron-Frobenius
operator of the same test-map as a classical propagator, 
and numerically demonstrate an analogous deviation from RMT of 
its eigenfunction statistics for both weak and strong couplings to a small opening in the phase space.    
Summary and conclusions are given
in section~\ref{sec:summary}. 

\section{Wavefunction intensity distribution}
\subsection{Theory}
\label{sec:supertheory}
Suppose $H_0$ is a GUE Hamiltonian. 
Since its eigenfunctions are complex valued, 
their intensities $x=|\langle a|\psi_0\rangle|^2$
at a certain state $|a\rangle$ 
%which we consider 
%localized in any space (e.g. coordinate, momentum,
%phase space) 
follow the exponential
distribution \cite{stoeck}
\begin{equation}
P(x) =
e^{-x}.
\label{eq:pt_cl}
\end{equation}             
Now we open the system at $|a\rangle$  \cite{Kap_opn} 
\begin{equation}
H = H_0 - i\frac{\Gamma}{2}|a\rangle\langle a|,
\label{eq:ham}
\end{equation}
and ask how the distribution 
of intensities $z=|\langle a|\psi\rangle|^2$
is changed with respect to the exponential,
when $\Gamma$ is small enough.
By  using perturbation theory
\cite{schomerus,polietal,polirap,newrussky},
we expand the amplitudes $\langle a|\psi\rangle$
in the first order as
\begin{equation}
\left<a|\psi_n\right> \simeq \left<a|\psi_n^{0}\right> -
i\Gamma\left<a|\psi_n^{0}\right>\sum_{p\neq n}\frac{|\left<\psi_p^{0}|a\right>|^2}
{2(E_n-E_p)}.
\label{eq:pt_ampl}
\end{equation}
Left and right eigenfunctions are in general distinct for the 
non-hermitian operator~(\ref{eq:ham}), but
they are just the complex conjugate of each other in first-order
perturbation regime.
We recognize two uncorrelated quantities, 
$\xi\equiv \left<a|\psi^0_n\right>$ whose real and imaginary parts are
Gaussian distributed
, and
$\eta\equiv \sum_{p\neq n}\frac{|\left<\psi_p^{0}|a\right>|^2}{2(E_n-E_p)}$,
following 
\begin{equation}
P_1(\eta) \propto \left(\frac{1}{1+\gamma^2\eta^2}\right)^{2}
\label{eq:py}
\end{equation}
with
$\gamma=\Delta E\pi^{-1}$, and 
$\Delta E$ average level spacing of $H_0$
(derivation in Appendix~\ref{app1} and~\cite{schomerus}).
We seek the distribution of the variable 
$z\equiv |\langle a|\psi\rangle|^2 =\xi^2+\Gamma^2\xi^2\eta^2$, namely 
\begin{eqnarray}
\nonumber
P(z) &=& \int d\xi d\eta\delta\left(z-\xi^2-\Gamma^2\xi^2\eta^2\right)P_0(\xi)P_1(\eta) \\
&=&
\frac{2\gamma}{\pi}
\int\frac{d\eta e^{-z/(1+\Gamma^2\eta^2)}}{(1+\gamma^2\eta^2)^2(1+\Gamma^2\eta^2)},
\label{eq:int_dist}
\end{eqnarray}
where $P_0(\xi)\propto e^{-\xi^2}$.
We immediately see that its expectation value
\begin{equation}
\langle z\rangle = \frac{\Gamma^2+\gamma^2}{\gamma^2}
\label{eq:int_dist_exp}
\end{equation}
always exceeds unity, meaning 
the opening produces a longer tail, and therefore a certain amount of localization
of the probability density takes place.  
\begin{figure}
\centerline{
\scalebox{1.8}{\includegraphics{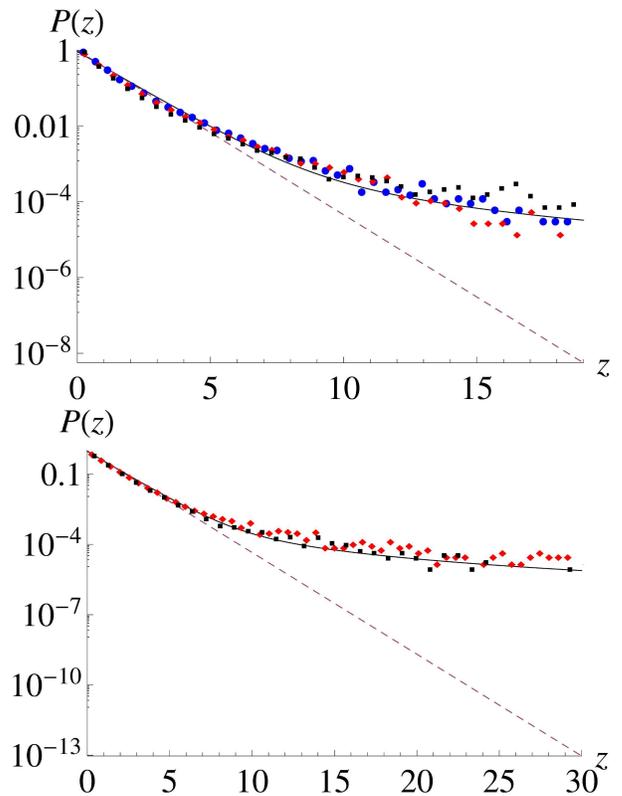}}}
%(a)\scalebox{.6}{\includegraphics{int_dist_gue.eps}}}
%\centerline{
%(b)\scalebox{.6}{\includegraphics{int_dist_cat.eps}}}
\caption{(a) Rescaled sample distributions of the overall wavefunction intensities
$P\left(z=|\left<a|\psi\right>|^2\right)$ in log scale, obtained
diagonalizing several realizations of a GUE Hamiltonian, for: 
$N=16384$ (dots, $8$ realizations), $N=4096$ (diamonds, $18$
realizations), $N=200$ (squares, $600$ realizations),  
and loss parameter $\Gamma=0.5$;
solid and dashed lines are the 
theoretical expectation~(\ref{eq:int_dist}) and the
exponential distribution~(\ref{eq:pt_cl}), respectively. (b) The same analysis with
the quantum cat map~(\ref{eq:incl_leaks}): $N=4096$ (diamonds, $28$ realizations),
$N=200$ (squares, $600$ realizations), and $\Gamma=1$.   
}
\label{distrib}
\end{figure}

\subsection{Numerical tests}
\label{sec:modello}
We now verify the theoretical intensity distribution~(\ref{eq:int_dist})
first by diagonalizing multiple realizations of the non-Hermitian Hamiltonian~(\ref{eq:ham}),
where both $H_0$ and the amplitudes $\left<a|\psi_n^{0}\right>$ 
are drawn from the Gaussian unitary ensemble (GUE).
The resulting probability distribution for the wavefunction intensities
$|\left<a|\psi\right>|^2$ in first-order perturbation regime
agrees with the expression~(\ref{eq:int_dist}) as shown in the example of
Fig.~\ref{distrib}(a). The dimension of the Hilbert space chosen 
ranges from $N=200$ to $N=16384$, suggesting that the result holds
in the semiclassical lmit. We will go back to this issue in sec.~\ref{sec:classic}.

Figure~\ref{distrib}(b) shows that our prediction for a perturbed GUE
Hamiltonian also fits the distribution of the wavefunction intensities of 
the quantized kicked cat map with a small opening. The 
classical evolution of the cat map reads~\cite{creagh,Cr_Lee}   
\begin{equation}
F_\epsilon = F_0 \circ M_\epsilon,
\label{eq:cat_map}
\end{equation}
with
\begin{equation}
F_0 : \left( \begin{array}{cc}
q' \\ p'
\end{array} \right) =
\left( \begin{array}{cc}
1 & 1 \\
1 & 2
\end{array} \right)
\left( \begin{array}{cc}
q \\ p
\end{array} \right)  \hspace{0.3cm} \text{mod} 1,
\label{eq:rotor}
\end{equation}
and
\begin{equation}
M_\epsilon :
\left( \begin{array}{cc}
q' \\ p'
\end{array} \right) =
\left( \begin{array}{cc}
q - \epsilon\sin(2\pi p) \\
p  \\
\end{array} \right) \hspace{0.3cm} \text{mod} 1
\label{eq:class_cat}
\end{equation}
The quantization of the map is given by \cite{creagh,han_ber}
\begin{equation}
U_\epsilon = U_0V_\epsilon
\label{eq:ueps}
\end{equation}
where
\begin{equation}
\left<q_j|U_0|q_k\right> = N^{-1/2}\text{e}^{i\pi/4}
\text{e}^{2\pi Ni(q_j^2-q_jq_k+q_k^2/2)}
\label{eq:uo}
\end{equation}
and
\begin{equation}
\left<q_j|V_\epsilon|q_k\right> = \sum_{p_m}
\frac{1}{N}\text{e}^{Ni\left(-\epsilon\cos2\pi p_m
+2\pi(q_j-q_k)p_m\right)}.
\label{eq:veps}
\end{equation}
The quantization of the linear map~(\ref{eq:rotor}) is known 
to possess pseudo-symmetries~\cite{KeatMez} that make the spectral 
statistics deviate from the Circular Unitary Ensemble (CUE), hence the use of the perturbation~(\ref{eq:class_cat})
to restore the RMT behavior. 
Here the opening is a minimum-uncertainty Gaussian wavepacket
\begin{equation}
\left<q|a\right> = \left(\frac{1}{\pi\hbar^2}\right)^{1/4}\text{e}^{-(q-q_0)^2/2\hbar+
ip_0(q-q_0)/\hbar},
\label{eq:gss_wvp}
\end{equation}
whose center $(q_0,p_0)$ is chosen at random on the unit 
torus (the scar at the origin~\cite{Cr_Lee,LRLK} is carefully avoided).
%~\footnote{we make sure the
%opening/probe state is chosen far enough from the scar 
%at the origin, known to significantly alter the wavefunction statistics 
%by itself~\ref{Cr_Lee}.}. 
The non-unitary propagator is realized 
by replacing $U_\epsilon$ of~(\ref{eq:ueps}) with 
\cite{Kap_opn,LRLK} 
\begin{equation}
U = \left(1 - \frac{\Gamma}{2}\left|a\right>\left<a\right|\right)U_\epsilon.
\label{eq:incl_leaks}
\end{equation}
All the steps of the derivation of Eq.~(\ref{eq:int_dist}) would still hold in this case, 
except for 
Eq.~(\ref{eq:py}), since the quasienergies of the cat map
follow the statistics of the CUE,
instead of the GUE's. Still, both are asymptotically equivalent
for $N\rightarrow\infty$~\cite{Mehta}. In our simulations we alternatively
set $N=200$ and $4096$, and produce an ensemble statistics of over $10^5$ states,          
by repeatedly diagonalizing 
the matrix~(\ref{eq:incl_leaks}) over different values of the kick strength
$\epsilon$, chosen at random within the range $[0.1,0.2]$.

\subsection{Strong coupling to the opening}
\label{sec:three}

%\subsection{Localization and resonance trapping}
%\label{sec:trap}

\begin{figure}
%(a)\scalebox{.6}{\includegraphics{int_vs_gamma.eps}}
%\vskip 0.5cm
\centerline{
%\scalebox{.42}{\includegraphics{kick_lscar_01.eps}}
\scalebox{.6}{\includegraphics{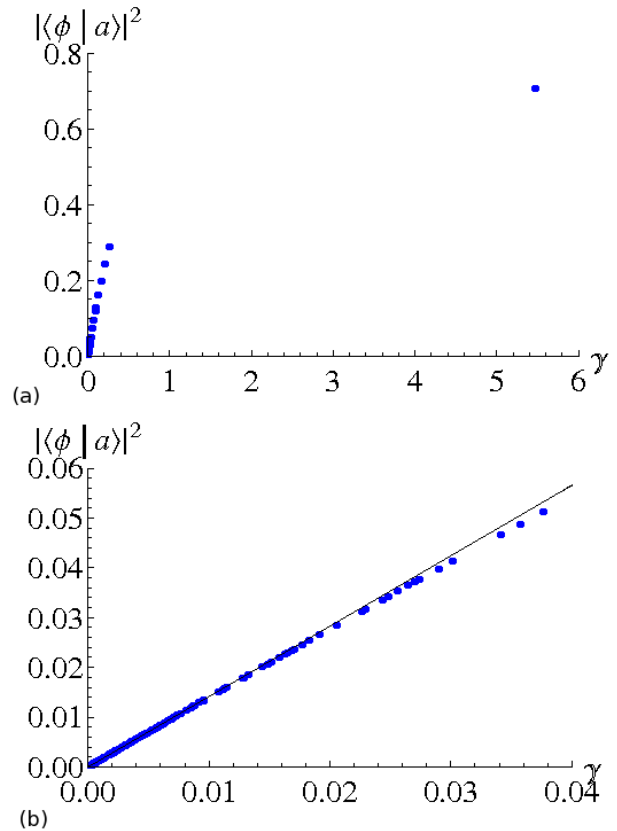}}}
%(b)\scalebox{.6}{\includegraphics{int_vs_gam_fit.eps}}}
\caption{(a) Overlaps between the opening and the left eigenfunctions 
of the quantized cat map vs. decay rates $\gamma$, showing the 
resonance trapping effect. (b) The linear part of the 
data is well described by Eq.~(\ref{BWL}), here $\Gamma=2\sqrt{2}$.}
\label{hloss_dist}
\end{figure}

When we further increase the coupling $\Gamma$ in the propagator~(\ref{eq:incl_leaks}), 
the curve~(\ref{eq:int_dist}) no longer fits the numerical data, as we leave the perturbation regime. 
A few short-lived left eigenstates are localized on the opening,
while the rest are characterized by 
intensity suppression together with small decay rates.
We will clarify the localization patterns of left 
and right eigenfunctions in section~\ref{sec:obs},
while we focus for the moment on the left ones.
%The picture is illustrated in  
%intensity $|\langle\phi|a\rangle|^2$
%versus escape rates $\gamma$ 
The overlaps between the open region and the eigenstates 
are presented as a function of the decay rate in 
Fig.~\ref{hloss_dist}(a),
reminiscent of the so-called `resonance trapping' 
effect 
%whose well-known results 
\cite{rotter91,kleinrot,russky,rotlet,RPPS,per_rot,isra,rotteralone,stockexp},
whose main results we summarize as follows.

%will be used in 
%Sec.~\ref{sec:model} 
%to better understand the  $|\langle\phi|a\rangle|^2$ vs. $\gamma$ dependence. 
%We also notice that in this regime
%the left and right eigenfunctions are well distinct.
%It turns out that the localization patterns induced by the
%opening differ most from left to right eigenfunctions 
%in the short-lived modes. We account for that  
%in Sec.~\ref{sec:obs}.  
%\begin{figure}
%\caption{
%$\Gamma=0.01$ (left), and 
%$\Gamma=0.2$ 
%(right)
%.}
%\label{trapping}
%\end{figure}
Consider the complex eigenvalues of $H$, $E_n-i\gamma_n$, $\gamma_n$ being the decay rates. 
It has been observed and explained~\cite{rotlet,russky} that when the overall loss 
$w=\sum_n\gamma_n$ is greater than the energy range $\Delta E$ where the levels are located,
there exists one particularly short-lived state $|\psi_1\rangle$, having decay rate
$\gamma_1=w-O\left(\Delta E/w\right)$, while the rest of the modes have 
$\gamma_{n\neq1}=O\left(\Delta E/w\right)$, so that they are `trapped' near the real axis, 
although still complex valued.
We will now use this property together with      
a $P-Q$ projection formalism
to explain the linear dependence of the
intensities $|\langle\phi_{n\neq1}|a\rangle|^2$ on the decay rates in this regime. 
Let $PHP$ be the projection
of the Hamiltonian onto the fast-decaying state, $P=|\psi_1\rangle\langle\phi_1|$,
and  $QHQ$ the projection on the remaining states,  
$Q=\sum_{n\neq1}|\psi_n\rangle\langle\phi_n|$. We first write an
eigenvalue of $QHQ$ as 
\begin{eqnarray}
\nonumber
(E_j-i\gamma_j) &=&
\langle\phi_j|QHQ|\psi_j\rangle 
\\
&=& \langle\phi_j|QH_0Q|\psi_j\rangle 
-i\frac{\Gamma}{2}\langle\phi_j|Q|a\rangle\langle a|Q|\psi_j\rangle.
\label{eq:towd_BW}
\end{eqnarray}
On the other hand, we know that $QHQ$ is
almost hermitian, 
so that, to a very good approximation, 
\begin{equation}
\left(Q|\psi_j\rangle\right)^\dagger = \langle\phi_j|Q.
\label{eq:return_first}
\end{equation}
We can now recognize the eigenvalues as
\begin{equation}
E_j  - i\gamma_j \simeq \langle\phi_j|H_0|\phi_j\rangle 
-i\frac{\Gamma}{2}|\langle a|\phi_j\rangle|^2,
\label{BW_rev}
\end{equation}
 where the first term is the expectation
value of a hermitian operator, hence a real number,
and therefore 
\beq
\gamma_j=\frac{\Gamma}{2}|\langle a|\phi_j\rangle|^2,
\ee{BWL}
so that we `return' to a Breit-Wigner kind-of law, as verified in Fig.~\ref{hloss_dist}(b)
for the simulations of the cat map.

\subsection{Left and right eigenfunctions}
\label{sec:obs}

\begin{figure}
\centerline{
\scalebox{2.85}{\includegraphics{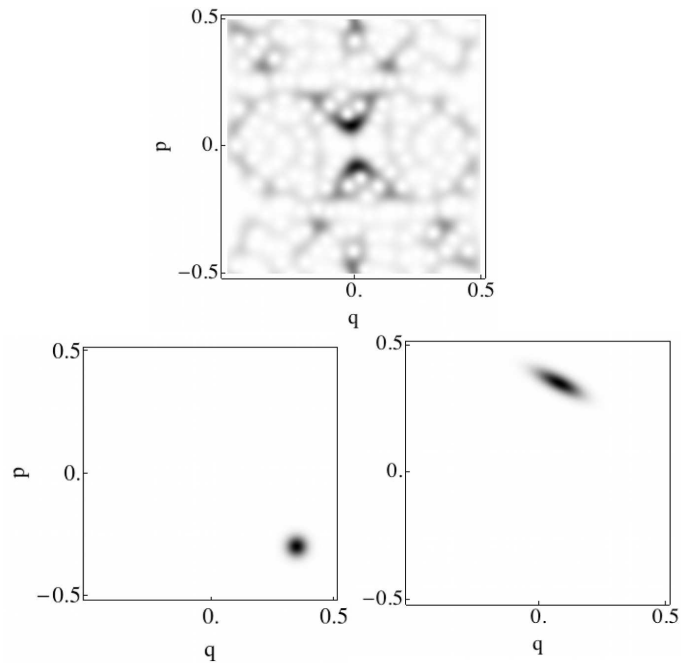}}}
%\scalebox{.35}{\includegraphics{master_mode.eps}}}
%\hskip 0.5cm
%\scalebox{.28}{\includegraphics{kick_fast_r.eps}}}
%\vskip 0.5cm
%\centerline{
%\scalebox{.35}{\includegraphics{left_shortlv.eps}}
%\scalebox{.35}{\includegraphics{right_shortlv.eps}}}
\caption{(Top) Husimi distribution of the fastest decaying
eigenstate of the
quantized cat map, and (bottom) 
the left and right 
eigenfunctions the same mode, when the map is open with
loss parameter $\Gamma=2\sqrt{2}$. 
The opening is placed exactly at the position where the Husimi distribution of
the left eigenfunction is localized.}
%The left 
%eigenfunction is supported exactly on the opening.} 
\label{left_right}
\end{figure}

We notice that in the strong-coupling regime the
left and right eigenfunctions of the propagator (\ref{eq:incl_leaks})
are well distinct.  
%and so must be their 
%the Husimi distributions of Fig.~\ref{left_right}, which we explain as follows:
In particular, we show in Fig.~\ref{left_right} the Husimi distributions
of the fastest-decaying eigenstates, whose
\textit{left} eigenfunctions only are supported on the opening. 
This is explained as follows.  

The discrete-time evolution operator~(\ref{eq:incl_leaks}) is
indeed split into unitary evolution $U_0$, and a projection 
describing the opening,
$P_o=1-|a\rangle\langle a|$, so that $U=P_oU_0$.
Given the eigenvalue $\lambda_j$
and its eigenfunctions  $\langle\phi_j|$ and $|\psi_j\rangle$, 
\begin{eqnarray}
\nonumber
\langle\phi_j|P_oU_0|x\rangle &=& \lambda_j^*\langle\phi_j|x\rangle \\
\langle x|P_oU_0|\psi_j\rangle &=& \lambda_j\langle x|\psi_j\rangle, 
\label{eq:ou_action}
\end{eqnarray}
the projection $P_o$ acts first on the 
left eigenfunction, so that in order for 
the loss to be maximal the
amplitudes $\langle\phi_j|x\rangle$ should be supported on
the opening, in our case the coherent state $|a\rangle$.
On the other hand, the unitary propagator $U_0$  
acts first on the \textit{right} eigenfunction $|\psi_j\rangle$:
in one time step we approximate the quantum evolution with the
classical map $F(x)$, and  
\begin{equation}
|\langle x|U_0|\psi_j\rangle|^2\simeq |F(\psi_j(x))|^2,
\label{one_step_qclas}
\end{equation}
%with $\rho_j=|\psi_j\rangle\langle\psi_j|$, 
so that  
loss/decay rate are highest if $|\psi_j\rangle$
is supported on the classical \textit{preimage} of the opening,
$F^{-1}(o)$ \cite{Schom_Tword,LRLK} (Fig.~\ref{left_right}). 
The localization patterns of left and right eigenfunctions will 
differ most when they occur where the
system is more sensitive to initial conditions, typically
away from fixed points
or stable/unstable manifolds of the classical map. 
%We provide an example in 
%Fig.~\ref{left_right}:
%for the same cat map in mode-trapping regime, 
%Husimi distributions of the shortest-lived 
%left and right eigenfunctions are portrayed. 
%As predicted by our argument, 
%the intensity is all concentrated
%on the opening in the left eigenfunction,
%and
%around its classical preimage in
%the right eigenfunction.    

In general the outcomes depend on how the 
propagation and the loss are arranged, 
which is usually 
%one we use 
$U=P_oU_0$ 
as in our model, 
but can be inverted 
sometimes \cite{fyod_somm}. 
%such as 
%in the discretization of the 
%\textit{classical} evolution operator, Eq.~(\ref{eq:ulam}). 
%In this case, the propagator can be split as  
%$L_o= L\cdot O$, so that applying the same reasoning
%as above, we find the fastest-decaying right eigenfucntions 
%to be localized on the opening, as seen in section~\ref{sec:classic},
%and the right ones concentrated on its image $f(O)$ (Fig.~\ref{left_right}).   

%Although an artifact of the one-step time-propagation of 
%both classical and quantum evolution operators, the effect 
%described above is inherent in quantum maps, 
%perhaps still present in other discretization schemes, 
%and for that reason,
%worth explaining. 
%In the light of the result of Eq.~(\ref{eq:ou_action}),
%we  
  
\section{Classical system}
\label{sec:classic}
In this section
we consider a classical chaotic map
with a small opening, again looking for 
deviations from RMT of the
sample distributions of the wavefunction intensities,
properly defined.  
%we provide a hint that the
%observed 
%we look for a deviation from RMT is a classical 
%effect. 
%To that aim we define and measure a classical equivalent to
%the wavefunction intensity.
%First, 
%We find the sample distributions of the wavefunction intensities, 
%constructed similarly with what is done in the closed and open quantum maps. 
The idea is to
fit the numerical data with analytic formulae obtained 
equivalently to~(\ref{eq:int_dist}) in perturbation regime, 
%distributions of $z$ numerically obtained in the classical cat map,
%as shown in Fig.~\ref{eig_stat}.
to then extend the analysis to a strongly-coupled opening,
as done in the quantum setting. 
%The result suggests that the change of the intensity
%distribution induced by the opening in the classical cat map may be at least qualitatively
%explained by the perturbation analysis done in the quantum setting. 
%Also, it indirectly shows that the localization observed for the open 
%quantum system in the weak-coupling regime has classical origin.

Using the density operator $\hat{\rho}$, the wavefunction intensities in the quantum 
regime can be written as
\begin{equation}
|\langle a|\psi\rangle|^2 =
\langle a|\hat{\rho}|a\rangle.
\label{eq:q_probe}
\end{equation}
%That I believe is correct, except that 
Here  $\hat{\rho}$ obeys the Liouville-von Neumann
equation \cite{Sakurai}
\begin{equation}
i\hbar\partial_t\hat{\rho} = [H,\hat{\rho}],
\label{vNeum}
\end{equation}
whose classical analog is \cite{goldie}
\begin{equation}
%\partial_t \rho  + \partial\cdot(\rho v) \,=\, 0
%\,.
\partial_t\rho = \{H,\rho\}.
\label{eq:liou}
\end{equation}

%The classical-to-quantum correspondence of our analysis
%stems from that observation.
%The classical Liouville equation is defined in phase space,
%not in coherent-state space, so that in general the  
%and thus the
%Wigner function
%\beq
%W(x,p) = \frac{1}{\pi} \int dy \psi^*(x-y/2)\psi(x+y/2)e^{-ipx/\hbar} %=
%N e^{-x^2/\sigma^2 -\sigma^2 p^2},
%\ee{DL:wigner}
%would be fitter than the Husimi to compare quantum with classical
%spectral properties.
%Wigner function evolves according to the equation
%\beq
%\partial_t W(x,p) =
%-\frac{p}{m}\partial_x W(x,p)
%+ \sum_{s=0}^\infty c_s(-\hbar^2)^s\partial_x^{2s+1}V(x)
%  \,\partial_p^{2s+1}W(x,p)
%\,.
%\ee{wigner_eq}

%However, in the classical limit a coherent state
%becomes a point in the phase space, and it can be shown that
%$P, Q$, and Wigner representations are all equivalent. In particular,
%Wigner equation~(\ref{wigner_eq}) reduces to the Liouville equation~(\ref{liouville}) (with a change of
%coordinates) when $\hbar\rightarrow 0$.

%Once the classical limit is taken,
%Liouville equation~(\ref{DL:wig_liou}) rules, and
The classical Liouville propagator can be written as
\beq
U^t_{cl} = e^{i\hat{L}t},
\ee{DL:lie}
where $\hat{L}=\{H,\cdot\}$ is the Liouville differential operator. In the Hamiltonian case 
$\hat{L}^\dagger=-\hat{L}$, and therefore the evolution~(\ref{DL:lie})
is unitary.
The classical evolution operator is supported on a space of generalized
functions, and its
the spectrum has a discrete and a continuous part
(Stone's theorem); all the eigenfrequencies lie on the unit circle. In particular,
ergodic and mixing systems only have one isolated eigenvalue, $e^{i\omega_0}=1$,  while the
rest of the spectrum is continuous~\cite{Gaspard}.

In reality every system experiences noise, coming for example from uncertainties or
roundoff errors. However small, noise breaks unitarity and changes the
spectrum of the Liouville propagator, from continuous to discrete~\cite{gasp95}. 
The (`leading')  unit eigenvalue 
is still there, but the rest of the spectrum moves inside the unit circle. 
In a closed system, the ground-state eigenfunction
of eigenvalue equal to unity (natural measure)
is real and positive definite, 
the density to which all initial conditions asymptotically converge.
The other eigenfunctions are in general complex and called `relaxation modes',
as they are associated with the decay of correlations~\cite{chaosbook}:
\beq
\langle g|\Lop|f\rangle = \sum e^{-\gamma_n}\langle g|\rho_n\rangle
\langle\tilde{\rho}_n| f\rangle
\ee{DL:corr_func}

The classical-to-quantum correspondence was studied by Fishman
and coworkers~\cite{fish,fish_clique}, 
who found that
the formal solution to the classical Liouville equation,
called Perron-Frobenius operator
[here $x=(q,p)$]
\beq
\left({\Lop}^t \circ \rho \right) (x) =
        \intM{\xInit} \prpgtr{x - \flow{t}{\xInit}} \rho(\xInit,0),
\ee{DL:PF}
when discretized, 
effectively behaves like the weakly noisy operator, and
has the same
spectrum as 
%the evolution operator
%$U_W$ 
the quantum propagator 
of the Wigner function  
in the classical limit
(a similar result was shown in~\cite{PaLuSr}).
%It was successively shown~\cite{fish_clique} that the effect of 
%such a discretization is equivalent to weak noise.
%Also, in 2000, Fishman and coworkers~\cite{fish_clique} showed that a discretization of
%the Perron-Frobenius operator for the kicked rotor affects the
%spectrum just like weak noise.

\begin{figure}
\centerline{
\scalebox{.45}{\includegraphics{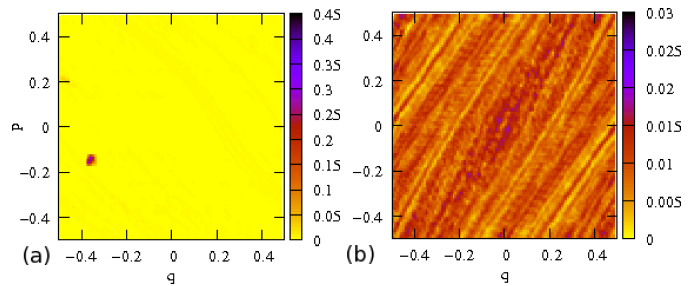}}}
%(a)\scalebox{.6}{\includegraphics{cat_dens1_hole.eps}}
%(b)\scalebox{.6}{\includegraphics{cat_dens13_hole.eps}}}
%\centerline{
%\scalebox{.5}{\includegraphics{opencol3168.eps}}
%\scalebox{.5}{\includegraphics{opencol3841.eps}}}
\caption{
Absolute values of (a) a fast- and (b) a slow-decaying eigenfunctions  
of the evolution operator~(\ref{DL:PF}) for the open cat map~(\ref{eq:cat_map}), with 
the hole located in the square $[-0.4,-0.3]\times[-0.2,-0.1]$, 
obtained diagonalizing a $10^4\times10^4$
discretization~(\ref{eq:ulam}). 
The manifold structure of the cat map is shown in (b).}
\label{open_eigs}
\end{figure} 

Based on that, we can say that
the noise introduced by the discretization
washes out the fine details of the chaotic dynamics, and
makes the random-wave assumption hold for    
the eigenfunctions of~(\ref{DL:PF}).   
These are complex valued (in the phase space), and therefore
their squared magnitudes (`intensities') $|\rho_n(x)|^2$ follow a 
$\chi^2$ distribution. 
%Strictly speaking
Ideally, the classical limit of  
the minimum-uncertainty wave packet would correspond to
just one cell of the phase-space discretization. Here we want to 
repeat the analysis carried in the quantum setting, and appreciate the
difference in the statistics of the eigenfunctions from the closed to the
open system. We believe this is done most effectively by taking 
the sum of the square magnitudes over a small phase-space interval, as
%their overlap with some region $\pS_o$ of the phase
%space:
\begin{equation}
\xi_n = \int_{\pS_o} dx_0\int dx |\rho_n(x)|^2\delta(x-x_0),
\label{eq:c_probe}
\end{equation}
that is the overlap of $|\rho_n(x)|^2$ with a 
delta function 
[classical limit of the coherent state $|a\rangle$ of Eq.~(\ref{eq:gss_wvp})] 
%for $\hbar\rightarrow 0$. 
supported on the probing region $\pS_o$. 
The quantum analog of Eq.~(\ref{eq:c_probe}) 
would be  $\sum_{a'}|\langle a'|\psi\rangle|^2$, over a number of probe
states. In that case, the
probability density $P\left(\sum_{a'}|\langle a'|\psi\rangle|^2\right)$
for the unperturbed system is a $\chi^2-$distribution with $M$ degrees of freedom,
\beq
P_M(\xi) = \frac{\xi^{M/2-1}e^{-\xi/2}}{\Gamma\left(\frac{1}{2}M\right)2^{M/2}},
\ee{chisq}
which becomes a Gaussian as $M\rightarrow\infty$. 
%(Here $\Gamma$ is the
%$\Gamma-$function).

\begin{figure}
\centerline{
\scalebox{.7}{\includegraphics{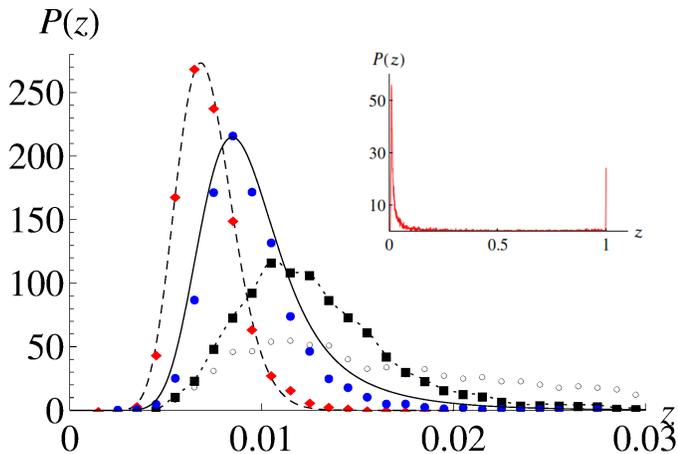}}}
\caption{Sample distributions of the intensities $\xi$ 
[given by Eq.~(\ref{eq:c_probe})]
of the eigenfunctions of the Perron-Frobenius operator
for the cat map~(\ref{eq:cat_map}):
(diamonds) closed system;
(filled dots) $25\%$ partial opening; (squares) $50\%$ partial opening;
(empty dots) $75\%$ partial opening; 
(dashed curve) Eq.~(\ref{chisq}) rescaled to the data set,
where the number of degrees of freedom
has been fitted from the data to $M=46$; (solid curve)
Eq.~(\ref{eq:ndfz}) with $M=46$, while $\tilde{\Gamma}=1$ is fitted from the data.
Inset: total opening at the same location; the peak at the tail represents
the instantaneous-decay states.} 
\label{eig_stat}
\end{figure}

We then perform numerical simulations on the classical cat map~(\ref{eq:cat_map}):
the Perron-Frobenius operator is discretized with Ulam method~\cite{ulam}
\begin{equation}
[\Lop]_{ij} \,=\, \frac{1}{|\pS_i|}
    \int_{\pS_i} \! dx  \int_{\pS_j} \!dy \,
    \delta(y-f(x))
\,.
\label{eq:ulam}
\end{equation}
The entries
$[\Lop]_{ij}$ are 
estimated using a straightforward Monte Carlo technique \cite{ErmShep},
based on counting how many trajectories starting from each $\pS_j$ land
in $\pS_i$.
%in both closed and open models.
A partial opening is realized by randomly decreasing the number of 
trajectories that start 
from the hole, which overall 
%In the open system, the hole 
covers a tiny 
$1\%$ 
of the available phase space. 
The $10^4\times10^4$ matrix~(\ref{eq:ulam}) is then diagonalized. 
Fig.~\ref{open_eigs}(a) shows a fast-decaying eigenfunction 
peaked in correspondence of the hole, as an extreme case of density enhancement
at the opening.     
We then measure the 
statistics of the intensities~(\ref{eq:c_probe}) in both the closed and open systems
(Fig.~\ref{eig_stat}):
while the sample taken from the closed system agrees with the law~(\ref{chisq}) 
($M$ is fitted from the data), the `intensities' on the opening exhibit a longer tail, like in the quantum regime.
We qualitatively account for this observation by performing the 
convolution~(\ref{eq:int_dist})  on 
the unperturbed distribution~(\ref{chisq}), 
this time in $M$ degrees of freedom, 
\beq
P(z) \propto \int r^{M-1}drd\eta\delta\left(z-r^2-\tilde{\Gamma}^2r^2\eta^2\right)e^{-r^2/2}P_M(\eta),
\ee{eq:class_start} 
where $r^2=\sum_{x}|\rho_n(x)|^2$ (here $x$ is discretized by our grid),
while the perturbation $\eta$ follows~\cite{schomerus}
\begin{equation}
P_M(\eta) \propto \left(\frac{1}{1+\gamma^2\eta^2}\right)^{1+M/2}.
\label{eq:pyn}
\end{equation} 
The outcome is
\beq
P_M(z) =  
C_M\int d\eta \frac{z^{M/2-1}e^{-z/2(1+\tilde{\Gamma}^2\eta^2)}}
{(1+\tilde{\Gamma}^2\eta^2)^{M/2}(1+\gamma^2\eta^2)^{1+M/2}}
\ee{eq:ndfz}
where $\gamma=\pi^{-1}$, 
$C_M=\frac{2^{-1-M/2}M}{\pi^{3/2}\Gamma\left(\frac{1+M}{2}\right)}$,
while $\tilde{\Gamma}$ is fitted from the sample distribution. 
Fig.~\ref{eig_stat} also shows 
two sample distributions of the 
intensities obtained 
for stronger couplings to the opening,
away from the perturbation regime: 
importantly, the trend of a flatter curve with 
a longer tail stays qualitatively the same, indicating 
an increasing number of fast-decaying states. 
A total opening introduces a number of 
instantaneous-decay states~\cite{Schom_Tword} that
completely localize on the hole. That generates
a peak at the very tail of the sample distribution,
whose shape remains otherwise qualitatively the same as 
for the partial openings (inset of Fig.~\ref{eig_stat}, note the scale). 
As seen, the quantum system in the same regime behaves differently, 
as the states that do not decay instantaneously are instead
long-lived, and the overall intensity distribution
is consistent with the resonance-trapping picture.        

We may now give an interpretation of our findings. 
An open system, be it classical or quantum, must allow for some 
fast-decaying initial conditions, among the others. 
Densities and wave functions must be expressible in terms of the eigenstates   
of the linear operators we are using. As a consequence, some of these eigenstates
also decay fast and are more concentrated on the opening and its preimages~\cite{Schom_Tword}. 
For weak coupling, both classical and quantum simulations fit this physical picture, and
behave likewise. 
Moreover, the calculated deviations of the intensity distributions from the 
RMT results all rely on perturbation theory,
which can be applied to any linear operator with a 
discrete, non-degenerate spectrum. That is the case for both the quantum Hamiltonian/propagator
and the discretized classical evolution operator.  

%Rather, 
%both time- and phase-space discretizations of the evolution
%are common to all     
On the other hand, classical and quantum systems behave differently
when strongly coupled to the opening,
the latter only displaying resonance 
trapping, while the former not showing any signatures of
mode interaction.

%\subsection{Physical interpretation}
%In the classical setting we pierce a hole in the phase space of $M$ 
%cells, so that the discretized Perron-Frobenius operator~(\ref{eq:ulam})
%has rank $N-M$, and there exist $M$ instantaneously-decaying states
%that must be supported on the opening. It was argued~\cite{Schom_Tword}
%that in reality there exist a lot more nontrivial short-lived states
%that escape ballistically after few (say $n$) iterations,
%and for that reason are mostly supported on the opening and its $n-th,(n-1)-th,...,1st$ 
%preimages. More precisely, the quantum map has 
%$N\left[1-exp(-\tau_{\mathrm{Ehr}}/\tau_{\mathrm{dwell}})\right]$. Although  
%Ehrenfest time is not defined in classical dynamics, the definition
%$\tau_{\mathrm{Ehr}}=\frac{1}{\lambda}\ln M$, provided in~\cite{Schom_Tword}
%can be regarded as a mixing time for the open chaotic map, that is the time 
%within which densities evolve `in chunks' before the stretching and
%folding tears them apart. In the case of a small opening and particularly 
%in our simulation, $\ln M\sim1$ and therefore $\tau_{\mathrm{Ehr}}$ is of 
%the order of the Lyapunov time. A rapid estimate yields that in the case
%of the classical map these non-trivial, fast-escaping states are about $2\%$
%of the total number of states, and constitute the very  
%tail of the distribution~(\ref{eq:ndfz}).

\section{Summary}
\label{sec:summary}
We have shown that:

$i)$ The overall wavefunction intensity distribution
of a random (GUE) Hamiltonian and a quantized chaotic map 
deviates from the predictions of RMT,
%in correspondence of  
%a single-channel, weakly coupled opening, 
when a weakly-coupled, single-channel opening is introduced.
The result holds in the semiclassical limit.
%where intensities are overall enhanced.
%In particular, intensities are enhanced in some eigenfunctions.
%produces a deviation from RMT of 
%at the location of the opening. 
%The pattern observed resembles the intensity
%distribution of scarred maps in closed systems, with some important differences regarding the
%relation between localization and decay of time correlations;

$ii)$ By further increasing the coupling to the open channel in our
model, few states localize on the
opening particularly strongly and decay fast, while 
the rest show the opposite
behavior: slow decay together with intensity
suppression at the opening. Using well-known results in the context
of the resonance trapping effect, 
we derived a linear relation between 
the intensities of the long-lived states and their 
decay rates, similar to Breit-Wigner law.  
In this framework, we also showed that
the difference in the localization  
patterns between fast-decaying
left and right eigenfunctions can be recognized as 
an artifact, inherent of the construction of 
open quantum maps.    
   
$iii)$ Analogous simulations of the discretized classical evolution
operator result in a deviation of the intensity distribution from
the RMT expectations akin to what is observed in the quantum setting, 
when the coupling to the opening is weak enough for perturbation 
theory to be valid. 
A stronger coupling to the opening increases the number of fast-decaying states,
so as to obtain a longer-tailed  intensity distribution, very different from the  
resonance trapping observed in the
quantum simulations. 

%In order to interpret our results, we  
%start from the important observation that
%RMT correctly describes the statistics of 
%the eigenfunctions of the classical Perron-Frobenius operator
%in the closed system. We believe that the
%discretization plays a major role in this process,
%effectively behaving like weak noise~\cite{fish_clique},
%that washes out the finest details of the chaotic dynamics,
%and makes the random-wave assumption hold in a classical setting.
%This result reminds of  the findings in~\cite{PaLuSr},
%where the entire quantum spectral autocorrelation
%of an open microwave billiard
%could be completely understood in terms of the classical Ruelle-Pollicott resonances.  

\section{Acknowledgements}
This research was
supported by Basic Science Research Program through
the National Research Foundation of Korea (NRF) funded by the 
Ministery of Science, ICT and future Planning
(2013R1A1A2011438). DL acknowledges support from the 
National Science Foundation of China (NSFC), 
International Young Scientists (11450110057-041323001).

\appendix

\section{Derivation of Eq.~(\ref{eq:py})}
\label{app1}
We start from the joint probability
distribution~\cite{schomerus} of 
$\eta=\sum_{p\neq n}\frac{|\xi|^2}{E_n-E_p}$
and $\zeta=\sum_{p\neq n}\frac{|\xi|^2}{(E_n-E_p)^2}$,
\begin{equation}
P(\eta,\zeta) \propto \frac{(1+\gamma^2\eta^2)^M}{\zeta^{2+3M/2}}
e^{-\frac{M\pi}{2\gamma\zeta}\left(1+\gamma^2\eta^2\right)},
\label{ab_dist}
\end{equation}
with $M$ number of degrees of freedom. We simply integrate over 
$\zeta$ to obtain the distribution of $\eta$, in one [Eq.~(\ref{eq:py})]
or $M$ [Eq.~(\ref{eq:pyn})] degrees of freedom.

\end{document}